\documentclass[twocolumn,pre,aps,showpacs,amsmath,amssymb]{revtex4-1}
\usepackage{graphicx}
\usepackage{bm}
\usepackage{epstopdf}

\newcommand{\ma}{\,+\,}
\newcommand{\me}{\,-\,}
\newcommand{\eq}{\,=\,}

\begin{document}

\title{Hamiltonian formulation towards minimization of viscous fluid fingering}
\author{Carlos Batista}
\email[]{carlosbatistas@df.ufpe.br}
\author{Eduardo O. Dias}
\email[]{eduardodias@df.ufpe.br}
\author{Jos\'e A. Miranda}
\email[]{jme@df.ufpe.br}
\affiliation{Departamento de F\'{\i}sica, Universidade Federal de Pernambuco,
Recife, Pernambuco  50670-901 Brazil}


\begin{abstract}
A variational approach has been recently employed to determine the ideal time-dependent injection rate $Q(t)$ that minimizes fingering formation when a fluid is injected in a Hele-Shaw cell filled with another fluid of much greater viscosity. However, such a calculation is approximate in nature, since it has been performed by assuming a high capillary number regime. In this work, we go one step further, and utilize a Hamiltonian formulation to obtain an analytical exact solution for $Q(t)$, now valid for arbitrary values of the capillary number. Moreover, this Hamiltonian scheme is applied to calculate the corresponding injection rate that minimizes fingering formation in a uniform three-dimensional porous media. An analysis of the improvement offered by these exact injection rate expressions in comparison with previous approximate results is also provided.
\end{abstract}
\maketitle

\section{Introduction}
\label{intro}

The viscous fingering instability~\cite{Saf} occurs when one fluid displaces another of higher viscosity in the effectively two-dimensional (2D)
environment of a Hele-Shaw cell, a device composed of two thinly separated parallel glass plates. Under such confined flow conditions, the
competition of surface tension, viscous forces, and pressure gradients on the two-fluid boundary induces the formation of peculiar interfacial
structures in the form of fingers (the so-called ``viscous fingers")~\cite{Rev,DeG,Par2,Saf86,Max,MW_rectang}. A popular version of the problem
considers flow in radial geometry, where a fluid of negligible viscosity is injected at constant injection rate against a viscous
fluid~\cite{Pat,Tho,Che,Cardoso,Mir4,Praud,Petri}. As time advances, these radially expanding fingers tend to split at their tips, developing
further into a complex branched morphology. Since the seminal work by Saffman and Taylor~\cite{Saf}, a large amount of literature has been dedicated
to understanding the formation of such beautiful fingering structures. In addition to its scientific
and academic importance, the study of viscous fingering is quite relevant to many industrial and technological applications ranging from flows in porous
media~\cite{Gorell}, enhanced petroleum recovery~\cite{Lake}, and microfluidics~\cite{Suh} to fluid mixing~\cite{Juanes}, chromatographic separation
processes~\cite{DeWit}, and biodynamics of cell fragmentation~\cite{Callan}.

In spite of the importance of the Hele-Shaw flow investigations mentioned above~\cite{Saf,Rev,DeG,Par2,Saf86,Max,MW_rectang,Pat,Tho,Che,Cardoso,Mir4,Praud,Petri},
where researchers focused on fingering formation and proliferation, significant progress has recently been made on a different facet
of the problem, namely the possibility of minimizing or suppressing the emergence of interfacial fingering. Some interesting strategies have been proposed
to contain the growth of viscous fingering patterns both in quasi-2D Hele-Shaw cells, as well as in 3D porous media. In these
studies~\cite{Cardoso,Stone,Talal13,Draga1,Draga2,Stone2,Draga3,Draga4,PRLZheng,Li2,Lesh,Mir11,Par,PRL,solo} the main purpose was
to try to avoid the uprising of interfacial undulations as much as possible. It is worth noting that the possibility of strongly restraining fingering
formation is of great practical interest under circumstances in which fingering growth is very undesirable like in oil recovery~\cite{Lake},
and in chromatographic separation procedures~\cite{DeWit}.

One first type of viscous fingering controlling protocol has been proposed in Refs.~\cite{Stone,Talal13,Draga1,Draga2,Stone2,Draga3,Draga4,PRLZheng},
where it has been demonstrated that viscous fingers can be considerably stabilized, by modifying the basic geometry of the classic Hele-Shaw cell setup,
while keeping constant the fluid injection rate $Q$ (area covered by unit time). For instance, Al-Housseiny {\it et. al}~\cite{Stone,Talal13} have shown 
that fingering formation can be properly inhibited if the upper cell plate is slightly tilted, so that the Hele-Shaw plates are no longer exactly parallel. 
A related stabilization scheme has been considered in Refs.~\cite{Draga1,Draga2,Stone2,Draga3,Draga4}, where the rigid upper cell plate is replaced by a flexible membrane.
More recently, Zheng and collaborators~\cite{PRLZheng} suggested a time-dependent control strategy in which fluid injection is applied while the Hele-Shaw
cell gap thickness is increased in time in the power-law form $b(t) \propto t^{1/7}$. In this situation, either the fingering instability is suppressed, or
a constant number of nonsplitting fingers are maintained during the fluid displacement process.

An alternative viscous fingering control and stabilization technique has been achieved in Refs.~\cite{Cardoso,Li2,Lesh,Mir11,Par,PRL,solo} just by manipulating the
injection rate, and keeping the traditional Hele-Shaw cell geometry unchanged. Meticulous experiments, analytical calculations, and sophisticated numerical
simulations have demonstrated that the development of the usual multibranched interfacial fingered morphology could be constrained by properly choosing the time
dependence of the injection flux $Q(t)$. For example, instead of employing a constant injection routine, Refs.~\cite{Cardoso,Li2,Lesh,Mir11} assumed a variant
injection rate which scaled with time like $Q(t) \propto t^{-1/3}$. Under such circumstances, the traditional ramified fingering patterns are suppressed, and
replaced by symmetric shapes containing a prescribed number of nonsplitting fingers.

In the spirit of Refs.~\cite{Cardoso,Li2,Lesh,Mir11}, it has also been verified that the use of a simple two-stage piecewise constant
injection process~\cite{Par}, in which a low injection rate is followed by stronger one, is able to restrain the establishment of interfacial
deformations. Finally, a suggestive variational method has been utilized to systematically search for a time-dependent injection rate $Q=Q(t)$ that
results in proper minimization of the viscous fingering instability~\cite{PRL}. As discussed in Ref.~\cite{PRL}, minimization of the linear interfacial 
amplitudes is equivalent to minimizing an integral, so that the search for a proper time-dependent injection rate $Q(t)$ for restraining the development 
of viscous fingering ends up being a variational problem. In this framing, the linear growth rate defines the Lagrangian of the system, and the ideal 
$Q(t)$ can be obtained by solving an Euler-Lagrange equation. It has been found that, in the limit of large capillary numbers (a measure of the relative 
strength of viscous and surface tension forces), the desired injection rate is remarkably simple, varying linearly with time $Q(t) \propto t$. 
This variational approach has been successfully applied to the related problem of a more complex 3D fluid flow in a uniform porous media~\cite{solo}. 
Again, for large capillary number conditions, it has been shown that the proper injection rate for minimizing viscous fluid fingering in porous media is also pretty simple, 
varying quadratically with time, i.e. $Q(t) \propto t^{2}$. Peculiarly, despite of their approximate nature (in the sense that the capillary number must be large), these Lagrangian solutions for $Q(t)$ are quite efficient in minimizing the amplitudes of the interfacial deformations. This Lagrangian controlling scheme has also been proved 
effective to damp interfacial perturbations if the displaced fluid is non-Newtonian~\cite{Ze1}, as well as for fluid flow displacements in curved Hele-Shaw 
cells~\cite{Ze2,Ze3}. As a matter of fact, it also works very well to restrain the development of interfacial disturbances in other pattern formation problems 
involving electric discharges and crystal growth~\cite{Ze4}.

In this work, we go one step further regarding the Lagrangian-based minimizing strategy proposed in Refs.~\cite{PRL,solo}, where {\it approximate solutions}
for the proper time-dependent injection rates $Q(t)$ to restrain viscous fluid fingering instability in Hele-Shaw cells and porous media have been found.
Here, instead of tackling the problems through a Lagrangian approach, we adopt a Hamiltonian formalism. It turns out that, by employing a Hamiltonian
scheme, we were able to find {\it exact solutions} (i.e., valid for arbitrary values of the capillary number) for the desired $Q(t)$ both in quasi-2D Hele-Shaw 
flows and in 3D porous media displacements. In this framework, comparisons between approximate and exact solutions of $Q(t)$ for both physical systems are 
provided, and the accuracy of the approximate Lagrangian approach presented in~\cite{PRL,solo} is discussed.

\section{Fingering minimization in Hele-Shaw cells}
\label{HSC}

\subsection{Hamiltonian formulation and exact solution for $Q(t)$}
\label{HSC2}

We consider a radial Hele-Shaw cell of gap spacing $b$, initially containing a viscous incompressible fluid of viscosity $\eta$. Then, a fluid of 
negligible viscosity is injected into the viscous fluid at injection rate $Q$ (equal to the area covered per unit time), which may depend 
on time. Notice that this is the most unstable viscosity-driven situation (maximum viscosity contrast case), which is the most challenging 
to control the development of fingering instabilities. Both fluids are Newtonian, and between them there exists a surface tension $\sigma$. 

We describe the perturbed fluid-fluid interface as 
\begin{equation}
\label{enfeite}
{\cal R}(\theta,t)= R(t) + \zeta(\theta,t), 
\end{equation}
where $\theta$ represents the azimuthal angle ($0 \le \theta \le 2 \pi$), and $R(t)$ is the time dependent unperturbed radius 
\begin{equation}
\label{R}
R(t)=\sqrt{R_{0}^{2} + \frac{1}{\pi} \int_{0}^{t} Q(t') dt'},
\end{equation}
with $R_{0}$ being the unperturbed radius at $t=0$.  In addition, 
\begin{equation}
\label{zeta1}
\zeta(\theta,t)=\sum_{n=-\infty}^{+\infty} \zeta_{n}(t) \exp{(i n \theta)}
\end{equation}
denotes the net interface perturbation with Fourier amplitudes $\zeta_{n}(t)$, and discrete azimuthal wave numbers $n$. Recall that our main goal 
is to find out what is the ``ideal" time-dependent injection rate $Q(t)$ for which interfacial perturbation amplitudes are as small as possible. This should be 
done by injecting a certain amount of the negligible viscosity fluid while keeping fixed initial [$R(t=0)=R_{0}$] and final [$R(t=t_{f})=R_{f}$] 
radii, where $t_{f}$ is the final time.

The variational method introduced in Ref.~\cite{PRL}, and later applied in Refs.~\cite{solo,Ze1,Ze2,Ze3,Ze4}, is based on a minimization process of the linear perturbation amplitude~\cite{Pat,Mir4}, yielding the {\it dimensionless} expressions
\begin{equation}
\label{relax} \zeta_{n}(t)=\zeta_{n}(0)~\exp \left\{I_{n}(R,{\dot R})\right\}
\end{equation}
with
\begin{equation}
\label{relax2}I_{n}(R,{\dot R})=  \int_{t_{0}=0}^{t_{f}=1} \lambda_{n}(R,{\dot R}) {\rm d}t,
\end{equation}
$\zeta_{n}(0)$ being the interfacial amplitude at initial time $t=t_0=0$, where
\begin{equation}
\label{GrowthRaten}
\lambda_{n}(R,\dot{R}) \eq \frac{\dot{R}}{R} (|n| \me 1) \me \frac{1}{{\rm Ca} R^3}\,|n|\,(n^2 \me 1)
\end{equation}
is the linear growth rate, and the overdot denotes total time derivative. Throughout this work we deal with dimensionless equations where 
length and time are rescaled by characteristics length $R_{f}$ and time $t_{f}$, respectively. Note that within this nondimensionalization scheme 
$R_{f}=1$ and $t_{f}=1$. Moreover,
\begin{equation}
\label{cap1}
{\rm Ca}=\frac{\eta U}{\sigma }\frac{R_{f}^{2}}{k}
\end{equation}
is a capillary number, with $U=R_{f}/t_{f}$ being a characteristic velocity, and $k=b^{2}/12$. Notice that a useful relationship connecting $R$, $\dot{R}$, and $Q$ can be readily extracted 
from Eq.~(\ref{R})
\begin{equation}
\label{Q}
Q(t)=2 \pi R \dot{R}.
\end{equation}

One interesting idea toward the search of the ideal $Q(t)$ [or equivalently, the ideal $R(t)$] has been put forward in Ref.~\cite{PRL}, where 
the amplitude of the mode with larger growth rate has been minimized. The general procedure is the following: first, one finds the value of 
$n$ that maximizes the growth rate~(\ref{GrowthRaten}),
$$  \left. \frac{\partial \lambda_n}{\partial n} \right|_{n = n_{\textrm{max}}}  \eq 0\; \Rightarrow \quad
  n_{\textrm{max}}  \eq  \sqrt{ \frac{1}{3} \,\left(\, 1\ma {\rm Ca} \dot{R} R^{2} \,\right)}.$$
Then, inserting this value of $n$ into Eq. (\ref{GrowthRaten}),  we obtain that the maximum growth rate is given by
\begin{equation}\label{Lag2D}
  \lambda_{ n_{\textrm{max}} }(R,\dot{R})  =  -\,\frac{\dot{R}}{R}  \ma  \frac{2\,\sqrt{3}\,}{9\,{\rm Ca}\,R^3} \left(  1 \ma {\rm Ca} \dot{R} R^2 \right)^{3/2}  .
\end{equation}
Therefore, due to Eqs. (\ref{relax})-(\ref{relax2}), we could say that, at the end of the injection process at $t_{f}=1$,  the relative amplitude of the 
mode with maximum growth rate is
\begin{equation}\label{Amplitude1}
  \left|\frac{\zeta_{ n_{\textrm{max}} }(1)}{\zeta_{ n_{\textrm{max}}}(0)}\right| \eq
\textrm{exp}\left[ \, \int_0^{1}\,\lambda_{ n_{\textrm{max}} }(R,\dot{R})\,dt \, \right] \,.
\end{equation}
As discussed in Refs.~\cite{PRL,solo,Ze1,Ze2,Ze3,Ze4}, this is not rigorously correct because $n_{\textrm{max}}$ evolves with time and, therefore, there is no such idea of single mode with maximum growth rate. Nevertheless, this is a useful and fruitful way to picture what is happening, as unequivocally demonstrated in Ref. \cite{PRL}. The relative amplitude (\ref{Amplitude1}) is minimized whenever the integral inside the exponential is minimized. Such variational problem can be solved by means the Euler-Lagrange Equation. Nonetheless, the equation of motion derived from the ``Lagrangian'' $\lambda_{n_{\textrm{max}}}(R,\dot{R})$ is nonlinear and, therefore, difficult to be solved exactly. In Ref. \cite{PRL}, it has been argued that in many cases of experimental interest the capillary number is sufficiently large, so that ${\rm Ca} |\dot{R} R^2| \gg 1$ and, therefore, it is acceptable to approximate the maximum growth rate (\ref{Lag2D}) by
$$ \lambda_{ n_{\textrm{max}} }(R,\dot{R})  \,\approx \,   -\,\frac{\dot{R}}{R}  \ma
\frac{2\,\sqrt{3} \dot{R}^{3/2} {\rm Ca}^{1/2}\,}{9}\,\,, $$
in which case the Euler-Lagrange equation yields a surprisingly simple differential equation $\ddot{R}\eq 0$, implying that $R(t)$ and $Q(t)$ [see Eq.~(\ref{Q})] 
are linear functions of time \cite{PRL}. However, it should be pointed out that there are other instances in which Hele-Shaw flows take place under considerably 
low capillary number circumstances (see for example Refs.~\cite{small1,small2,small3}, and references therein). 

The aim of the present section is to go further and obtain the function $R(t)$ that minimizes the amplitude with maximum growth rate {\it without} 
assuming that the capillary number must be large. This amounts to solving a nonlinear differential equation. To accomplish this, we shall simplify such 
nonlinear problem by using the fact that our Lagrangian $\lambda_{n_{\textrm{max}}}$ does not depend explicitly 
on the time and, therefore, its associated Hamiltonian is conserved.

The Hamiltonian associated to the full Lagrangian written out in Eq. (\ref{Lag2D})  is
\begin{align*}
   H_{ n_{\textrm{max}}} \eq &  \dot{R}\,\frac{\partial \lambda_{ n_{\textrm{max}}} }{\partial \dot{R}}  \me \lambda_{ n_{\textrm{max}}}  \\
  \eq & \frac{R^4\,\dot{R}^2 \me (R^2\,\dot{R} / {\rm Ca}) \me (2 / {\rm Ca}^{2})  }{ 3\,\sqrt{3/ {\rm Ca}} \, R^3\, \sqrt{ (1/{\rm Ca}) \ma R^2\,\dot{R} \ } }\,.
\end{align*}
Since the Hamiltonian is a constant of motion, it follows that the above expression is a constant. For future convenience, we shall set such a constant as 
$(c_{1} \sqrt{{\rm Ca}})/9$, so that
\begin{equation}
\label{HamiltEq}
   \frac{R^4\,\dot{R}^2 \me (R^2\,\dot{R} / {\rm Ca}) \me (2 / {\rm Ca}^{2})  }{R^3\, \sqrt{ (1/{\rm Ca}) \ma R^2\,\dot{R} \ } }\, \eq \frac{c_1}{\sqrt{3}} \,.
\end{equation}
Equation~(\ref{HamiltEq}) constitutes a first order differential equation to be solved for $R(t)$, with the constant $c_1$ being determined by the initial conditions $R(0)$ and $\dot{R}(0)$  or, equivalently, by the boundary conditions $R(0)$ and $R(t_f)$. The use of the Hamiltonian amounts to an important technical improvement compared with the Lagrangian approach, inasmuch as the latter path would lead to a second order differential equation, which generally is more difficult to be solved. 

At this point, our job is to find the general solution for the nonlinear first order differential equation (\ref{HamiltEq}). To succeed in doing this, it is useful to define the 
function $f(t)\eq [R(t)]^3$, in terms of which Eq. (\ref{HamiltEq}) is written as
\begin{equation}\label{Hamiltf}
  [ \dot{f} \me (6 / {\rm Ca})]\, \sqrt{\dot{f} \ma (3 / {\rm Ca})} \eq 3\, c_1  \,  f \,.
\end{equation}
Then, differentiating this expression with respect to time, and defining the function $F(t) = \sqrt{\dot{f} + (3 / {\rm Ca})}$, we obtain the following 
differential equation
$$ [ F^2 \me (3 / {\rm Ca})  ]\,( \dot{F} \me c_1 ) \eq 0 \,. $$
Therefore, either $F\eq \pm \sqrt{3 / {\rm Ca}}$ or $\dot{F}$ is equal to the constant $c_1$. In the first case, we have that $\dot{f}\eq 0$, so that $R(t)$ would be constant, which is not the dynamical solution that we are looking for. Thus, we conclude that $\dot{F}$ must be some constant, which means that
$$ F(t) \eq c_1\,t \ma c_2 \,, $$
where $c_1$ and $c_2$ are constants.  Once we have found $F(t)$, it is a simple matter to obtain $f(t)$,
\begin{align}
   \sqrt{\dot{f} \ma (3 / {\rm Ca}) } \eq F\eq c_1\,t \ma c_2 \quad \Rightarrow \quad  \nonumber \\
    f(t) \eq \frac{t}{3}\, [c_1^2\,t^2 \ma 3\,c_1\,c_2\,t \ma 3\,c_2^2 \me (9 / {\rm Ca})] \ma c_3 \,, \label{f3c}
\end{align}
where $c_3$ is another integration constant. However, we have obtained this solution for $f(t)$ by means of differentiating Eq. (\ref{Hamiltf}). Therefore, we should check whether the latter solution satisfies Eq. (\ref{Hamiltf}). Actually, we can already anticipate that, in general, the above expression for $f(t)$ is not in accordance with 
Eq.~(\ref{Hamiltf}). Indeed, Eq.~(\ref{Hamiltf}) is a first order differential equation and, therefore, its solution should have one integration constant which together with $c_1$  sums up a total of two free constants in the general solution. However, the solution presented in Eq. (\ref{f3c}) has three arbitrary constants. Consequently, we conclude that in order for the solution $f(t)$ presented in (\ref{f3c}) to satisfy the relation (\ref{Hamiltf}), the constants $c_1$, $c_2$ and $c_3$ should not be all mutually independent. Indeed, by substituting the expression for $f(t)$ in Eq.~(\ref{f3c}) into Eq.~(\ref{Hamiltf}), one can check that $c_3$ is given by
$$ c_3 \eq \frac{c_2^{\,3} \me (9 / {\rm Ca}) \,c_2\ }{3\,c_1} \,,$$
in which case the solution for $f(t)$ is the following
$$ f(t) \eq \frac{ (c_1\,t \ma c_2)^3\,\me (9 / {\rm Ca}) \,(c_1\,t \ma c_2) }{3\,c_1} \,. $$
Since the function $f$ was defined to be such that $f(t)\eq [R(t)]^3$, we finally conclude that
\begin{equation}\label{R2D}
  R(t) \eq  \left[\, \frac{ (c_1\,t \ma c_2)^3\,\me (9 / {\rm Ca}) \,(c_1\,t \ma c_2) }{3\,c_1} \,\right]^{1/3} \,,
\end{equation}
where $c_1$ and $c_2$ are constants that are determined by the boundary conditions $R(0)=R_{0}$ and $R(t_f)=R_{f}$. For example, in the 
large capillary number limit (${\rm Ca} \gg 1$), it is immediate to verify that (recall that within our nondimensional scheme $R_{f}=1$, and $t_{f}=1$)
\begin{align*}
 c_1 &= \frac{\sqrt{3}\, \left[\,R_{f} - R_{0}\,\right]^{3/2}}{t_f^{3/2}} \,= \sqrt{3}\, \left[\,1 - R_{0}\,\right]^{3/2}, \\
 c_2 &= \frac{\sqrt{3}\,R_{0}\, \left[\,R_{f} - R_{0}\,\right]^{1/2}}{t_f^{1/2}} \,= \sqrt{3}\,R_{0}\, \left[\,1 - R_{0}\,\right]^{1/2},
\end{align*}
so that $R(t)$ becomes a linear function of time.
In the general case of arbitrary values for the capillary number, it is also simple to find the integration constants $c_1$ and $c_2$, but in such a 
case this is tantamount to finding the roots of a polynomial of order three. Therefore, once we have found $c_1$, let us say, the constant $c_2$ will 
admit 3 solutions, two of which will be nonphysical. For instance, two solutions for $c_2$ can be complex. 

Now, inserting the solution (\ref{R2D}) into Eq. (\ref{Q}), we obtain that the ideal injection rate is
\begin{eqnarray}
\label{Q2D}
Q(t) \eq \left ( \frac{c_{1}}{9} \right )^{1/3} \, \frac{ 2 \pi [ (c_1\,t \ma c_2)^2\,\me (3 / {\rm Ca})] }{  \left[(c_1\,t \ma c_2)^3\,\me (9 / {\rm Ca}) \,(c_1\,t \ma c_2)\right]^{1/3}}. \nonumber \\
\end{eqnarray}
Equation~(\ref{Q2D}) is one of the central results of this work, offering a closed form expression, valid for arbitrary values of ${\rm Ca}$, for the time-dependent injection rate that minimizes the interfacial perturbation amplitudes in the radial Hele-Shaw problem. If we take the large capillary number limit of this expression, we find that $Q(t)$ is a linear function of the time, in accordance with the approximate result obtained in Ref. \cite{PRL} [see their Eq. (7)]. By expanding the solution (\ref{Q2D}) in a power series of $(1 / {\rm Ca})$, we conclude that the solution obtained in the mentioned reference comprises just the zeroth order part of the series. Therefore, from the theoretical point of view, the results obtained in the present paper represent a considerable improvement. In what follows, we will analyze how this improvement reflects on the minimization of the viscous fingering process in Hele-Shaw cells.

\subsection{Comparing the exact Hamiltonian solution with the approximate Lagrangian solution}
\label{HSC3}

\begin{figure}[b]
\includegraphics[width=3.0 in]{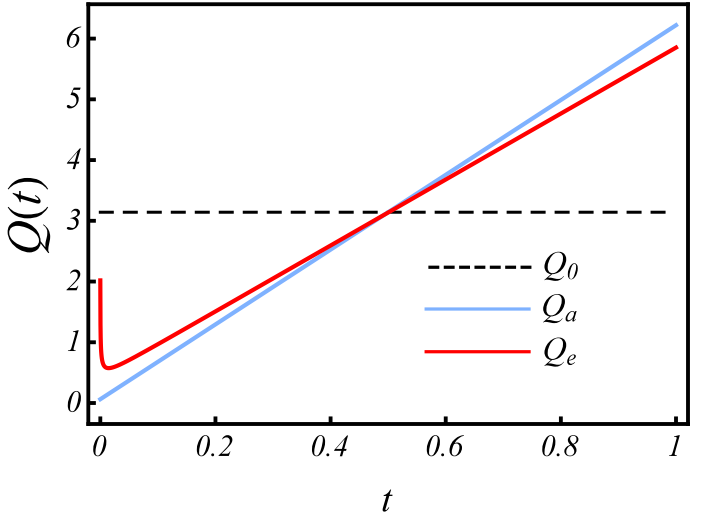}
\caption{(Color online) Sketch of the injection rate as a function of time, for the exact injection rate $Q_{e} (t)$ given by 
Eq.~(\ref{Q2D}), for the approximate injection rate $Q_{a}(t)=c_{1} + c_{2}t$ obtained in Ref.~\cite{PRL} [see their Eq. (7)], and 
for the constant injection rate $Q_{0}$ given by Eq.~(\ref{constant}). Notice that the total volume of injected fluid (area under 
the curves) in the interval $0 \le t \le t_{f}$ is the same for all these pumping rate cases.}
\label{injection}
\end{figure}

In this section, we compare the stabilization process of the perturbation amplitudes provided by the Hamiltonian exact solution given by Eq.~(\ref{Q2D}) 
[hereafter denoted by $Q_{e}(t)$], with the equivalent process offered by the approximate injection rate solution $Q_{a}(t)=c_{1} + c_{2}t$, originally 
calculated in Ref.~\cite{PRL}. In addition, we contrast these two amplitude minimization protocols, with the usual constant injection rate 
procedure~\cite{Saf,Rev,DeG,Par2,Saf86,Max,MW_rectang,Pat,Tho,Che,Cardoso,Mir4,Praud,Petri}, which considers the insertion of a specific volume of fluid 
at a time-independent injection rate $Q_{0}$, and results in the development of deformed interfacial structures. By using the dimensionless version 
of Eq.~(\ref{R}), with $R_{f}=1$ and $t_{f}=1$, one readily obtains that 
\begin{equation}
\label{constant} 
Q_{0}=\pi(1 - R_{0}^{2}).
\end{equation}

In Fig.~\ref{injection}, we plot the behavior of $Q_0$, $Q_a(t)$ and $Q_e(t)$ as time progresses, for capillary number 
${\rm Ca}=625$, and initial radius $R_0=0.01$. These values of ${\rm Ca}$ and $R_{0}$ are consistent with the physical 
parameters used in typical experimental realizations of Hele-Shaw flows~\cite{Saf,Rev,DeG,Par2,Saf86,Max,MW_rectang,Pat,Tho,Che,Cardoso,Mir4,Praud,Petri}. 
We begin by calling the readers' attention to the fact that the area under each curve in Fig.~\ref{injection} has the same magnitude, so that 
an equal amount of fluid is injected at the end of the pumping process, for the three injection schemes. By inspecting Fig.~\ref{injection}, we verify that 
the general behaviors of $Q_a(t)$ and $Q_e(t)$ are not very different. Only at the beginning of the injection process, one notices that there exists a 
more significant difference between the behaviors of $Q_a(t)$ and $Q_e(t)$. This happens due to the assumption ${\rm Ca}R^2 \dot R \gg 1$ 
considered for the approximate case~\cite{PRL}. When $t \rightarrow 0$, we have $R \rightarrow R_0$, so if $R_0$ is relatively small, this approximation 
no longer holds. Consequently, $Q_a(t)$ and $Q_e(t)$ must have different time evolutions at the beginning of the injection process, as clearly illustrated 
in Fig.~\ref{injection}.

\begin{figure}
\includegraphics[width=3.2 in]{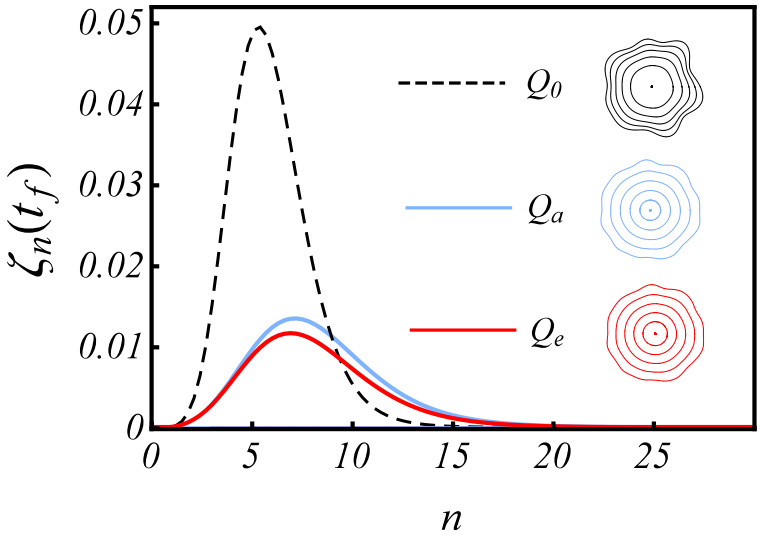}
\caption{(Color online) Plot of the interfacial perturbation amplitudes $\zeta_{n}(t)$ [as given by~Eq.~(\ref{relax})], as a function
of Fourier mode $n$ at $t=t_f$, when the three injection rates are used: $Q_0$, $Q_a(t)$ and $Q_e(t)$. The resulting interfaces obtained 
for each of these injection rate schemes are also shown.} 
\label{bands}
\end{figure}

Now, with the help of Fig.~\ref{bands}, let us compare the resulting interface perturbation amplitudes
at $t=t_{f}$, obtained by using $Q_a(t)$, $Q_e(t)$, and $Q_{0}$. We point out that the results presented 
in Fig.~\ref{bands} are obtained by utilizing the same physical parameters as those used in 
Fig.~\ref{injection} [i.e., $R_0=0.01$, ${\rm Ca}=625$, and $\zeta_{n}(0)=R_{0}/65$]. Figure~\ref{bands} plots the perturbation amplitudes 
given by~Eq.~(\ref{relax}) at $t_f$, for the approximate injection rate case $\zeta_{n}^{a}(t_f)$, the
exact ideal pumping rate situation $\zeta_{n}^{e}(t_f)$, and for the equivalent constant injection rate case 
$\zeta_{n}^{0}(t_f)$, as functions of the Fourier mode $n$. By examining Fig.~\ref{bands}, one can
see a substantial reduction of the final perturbation amplitudes when both time-dependent 
injections $Q_a(t)$ and $Q_e(t)$ are used. Moreover, it is also evident that an improved reduction 
of the perturbation amplitudes is offered by the exact injection rate $Q_e(t)$. This indicates that
the exact solution [Eq.~(\ref{Q2D})] actually does a better job in minimizing the strength of the 
interfacial perturbation amplitudes. 

On the right side of Fig.~\ref{bands}, we depict the time evolution leading to the final shape of the fluid-fluid interface 
at $t=t_f$, for the three injection rates. The resulting interfaces correspond to the amplitudes shown on left panel of 
Fig.~\ref{bands}. The patterns for each final interface have the same initial conditions (including
the random phases attributed to each mode), and 15 Fourier modes have been considered in the linear calculation. It is apparent that fingering 
formation is considerably inhibited when both ideal injections $Q_a(t)$ and $Q_e(t)$ are used. However, it is also 
worthwhile to note that, at the linear level, it is hard to observe any dramatic morphological changes between the 
final interfacial shapes for $Q_a(t)$ and $Q_e(t)$ cases. 

\begin{figure}
\includegraphics[width=3.2 in]{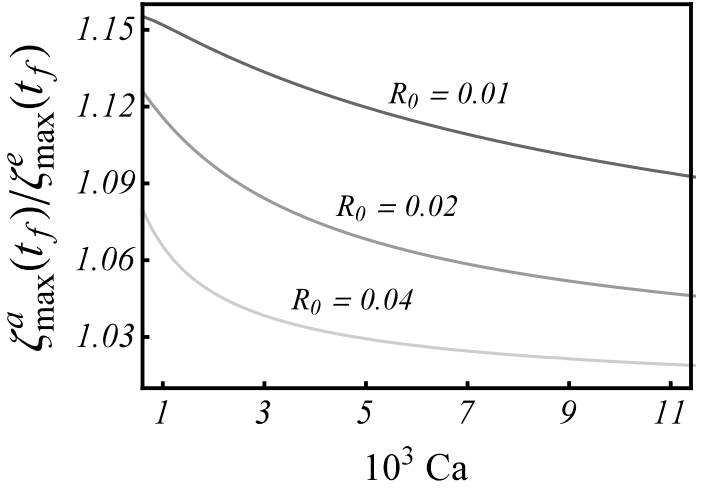}
\caption{Amplitude ratio $\zeta^a_{{\rm max}}(t_f)/\zeta^e_{{\rm max}}(t_f)$ as a function of
${\rm Ca}$, for $R_0=0.01, 0.02$, and 0.04. Here  $\zeta^a_{{\rm max}}(t_f)$ [$\zeta^e_{{\rm max}}(t_f)$]
denotes the maximum amplitude for approximate [exact] injection at
$t=t_f$.} \label{sims}
\end{figure}

It is important to analyze the behavior of the exact 
injection process when the capillary number ${\rm Ca}$ is varied. 
Regarding this point, Fig.~\ref{sims} plots the maximum amplitude for the approximate
pumping situation divided by the maximum amplitude calculated by
using the exact injection rate [$\zeta^a_{{\rm max}}(t_f)/\zeta^e_{{\rm max}}(t_f)$], 
as a function of ${\rm Ca}$, at final time $t=t_f$. 
We consider three values of $R_0$: $0.01$, $0.02$, and $0.04$. From this figure, 
it is clear that the ratio $\zeta^a_{{\rm max}}(t_f)/\zeta^e_{{\rm max}}(t_f)$ decays when 
${\rm Ca}$ is increased, and when larger values of $R_0$ are considered. 
In addition, Fig.~\ref{sims} shows that the amplitudes of the perturbations for
$Q=Q_e(t)$ are guaranteed to be smaller than the ones obtained by
the approximate injection process. Moreover, we observe that
for small values of ${\rm Ca}$ and $R_0$, $Q_a(t)$ can promote a
final perturbation amplitude $15\%$ larger than the exact injection
process $Q_e(t)$. This observation reinforces the idea that the 
exact solution obtained in this work [Eq.~(\ref{Q2D})] can provide a considerable improvement in the interfacial 
amplitude minimization process, as compared to the performance associated to 
the approximate solution case studied in Ref.~\cite{PRL}. This behavior can be explained by analyzing the
approximation used to obtain $Q_a(t)$~\cite{PRL}. Recall that $Q_a(t)$ is
calculated by considering that ${\rm Ca}R^2\dot R \gg 1$. Therefore, for lower 
values of the capillary numbers ${\rm Ca}$ and initial radii $R_0$, this 
approximation is no longer valid. In these circumstances, $Q_a(t)$ generates a less efficient stabilization 
of the perturbation amplitudes. This is observed in Fig.~\ref{sims}, where one can 
verify that the largest discrepancy between $Q_e(t)$ and $Q_a(t)$ occurs for smaller 
values of ${\rm Ca}$ and $R_0$.

\section{Fingering minimization in 3D porous media}
\label{porous1}

\subsection{Hamiltonian formulation and exact solution for $Q(t)$}
\label{porous2}

In this section we consider the fingering process when a fluid of negligible viscosity is injected through a punctual source in a 3D uniform porous 
media that is initially filled with a Newtonian viscous fluid of viscosity $\eta$. In this case the Saffman-Taylor instability takes place, giving rise to fingers that deform 
the initial spherical shape of the interface that separates the two fluids. Just like in the case of the Hele-Shaw flow treated in Sec.~\ref{HSC2}, we would like 
to find the time-dependent injection rate $Q(t)$ that minimizes the formation of fingers.

The evolving 2D surface that separates the fluids is defined by the equation $r = \mathcal{R}(t,\theta,\phi)$, where the coordinates $r$, $\theta$ and $\phi$ are spherical coordinates whose center $r=0$ is the point of injection. In analogy with the previous section, it is advantageous  to write the function $\mathcal{R}(t,\theta,\phi)$ as the sum of an angle independent part $R(t)$ that represents the radius of the interface if no fingering had occurred, plus a function $\zeta(t,\theta,\phi)$ that measures the size of the fingers
$$ \mathcal{R}(t,\theta,\phi) \eq R(t) \ma  \zeta(t,\theta,\phi) \,,$$
where $\theta$ is the polar angle ($0 \le \theta \le \pi$), $\phi$ is the azimuthal angle ($0 \le \phi \le 2 \pi$), and 
\begin{equation}
\label{R2}
R(t)=\left [ R_{0}^{3} + \frac{3}{4 \pi} \int_{0}^{t} Q(t') dt'\right ]^{1/3}.
\end{equation}
It is also convenient to expand the interface perturbation $\zeta$ in the basis of spherical harmonics
$$ \zeta(t,\theta,\phi) \eq \sum_{\ell=1}^{\infty} \, \sum_{m = -\ell}^{\ell} \,  \zeta_{\ell m}(t)\, Y_{\ell m}(\theta, \phi) \,.$$
Adopting the well-established  model of Chuoke \textit{et al.} for porous media \cite{Art2}, and using the linear 
stability analysis presented in Ref.~\cite{solo}, one finds that, up to the first order in $\zeta/R$, the dynamics of the perturbation 
amplitude $\zeta_{\ell m}$ is governed by the following equation
\begin{equation}\label{dZeta3D}
 \frac{d}{dt}\,\zeta_{\ell m} \eq \lambda_{\ell}(R,\dot{R})\, \zeta_{\ell m} \,
\end{equation}
where the dimensionless linear growth rate $\lambda_{\ell}$ is given by
\begin{equation}\label{Lagl}
  \lambda_{\ell}(R,\dot{R}) \eq \frac{ \dot{R} }{R} \, (\ell -1) \me \frac{1}{{\rm Ca}R^3}\,(\ell + 2)\,(\ell^2 - 1) \,
\end{equation}
with ${\rm Ca}$ is the capillary number as defined in Eq.~(\ref{cap1}), and now $k$ denotes the permeability of the porous medium, while $\sigma$ is an 
effective surface tension of the system~\cite{Art2}.

The mode of largest growth rate is the one with $\ell\eq \ell_{\textrm{max}}$, where
$$  \left. \frac{\partial \lambda_\ell}{\partial \ell} \right|_{\ell=\ell_{\textrm{max}}} \eq  0 \; \Rightarrow \quad
  \ell_{\textrm{max}}  \eq  \frac{ \sqrt{3\,{\rm Ca}\,R^2\,\dot{R} \ma 7 } }{3} \me \frac{2}{3} \,.  $$
Inserting this value into Eq.(\ref{Lagl}), we obtain that the maximum growth rate is
\begin{eqnarray}
\lambda_{\ell_{\textrm{max}}}(R,\dot{R}) &\eq& -\, \frac{ 5\,\dot{R} }{3\, R} \nonumber \\ 
                                         &\ma& \frac{2}{27 {\rm Ca} R^3}\,\left[\,3\,{\rm Ca}\,R^2\,\dot{R} \ma 7 \right]^{3/2}  \nonumber\\
                                         &\ma& \frac{20}{27 {\rm Ca} R^3} \,. \nonumber \\
\label{Lag3D}
\end{eqnarray}
Then, by integrating Eq. (\ref{dZeta3D}), one can say that the relative amplitude of the mode with maximum growth rate is
\begin{equation}\label{Amplitude2}
  \left|\frac{\zeta_{ \ell_{\textrm{max}} m}(1)}{\zeta_{ \ell_{\textrm{max}}m}(0)}\right| \eq
\textrm{exp}\left[ \, \int_0^{1}\,\lambda_{ \ell_{\textrm{max}} }(R,\dot{R})\,dt \, \right] \,.
\end{equation}
Similarly to what happened in Sec.~\ref{HSC2}, the minimization of this amplitude occurs when the integral inside the 
exponential in Eq. (\ref{Amplitude2}) assumes the minimum value, which is a variational problem with $\lambda_{ \ell_{\textrm{max}} }$ playing the role of the 
Lagrangian. Although flow in porous media can occur at low capillary numbers (see Refs.~\cite{Rev,Gorell,small4}, and references therein), in Ref. \cite{solo} it has been 
argued that in many experimental setups one has that ${\rm Ca} | R^2 \dot{R}| \gg 1$ (see also Ref.  \cite{Art2}), so that the Lagrangian (\ref{Lag3D}) can be rewritten as
$$   \lambda_{\ell_{\textrm{max}}}(R,\dot{R}) \,\approx \, -\, \frac{ 5\,\dot{R} }{3\, R} \ma \frac{2\,\sqrt{3 {\rm Ca}}}{9}\,\dot{R}^{3/2}  \,,$$
in which case the Euler-Lagrange equation is $\ddot{R}=0$, namely the ideal $R(t)$ is a linear function of the time. Moreover, from Eq.~(\ref{R2}) one can see 
that 
\begin{equation}
\label{newQ}
Q(t)=4 \pi R^{2} {\dot R}, 
\end{equation}
resulting in an ideal injection rate that varies quadratically with time.

In this section we improve the result of Ref. \cite{solo} by solving the variational problem for the full Lagrangian (\ref{Lag3D}), \textit{i.e.}, without 
assuming that ${\rm Ca}$ must be large. In order to accomplish this goal, we use the fact that the Hamiltonian associated to the Lagrangian 
$\lambda_{\ell_{\textrm{max}}}$ is constant, in view of the fact that $\lambda_{\ell_{\textrm{max}}}$ does not depend explicitly on the time. This approach will be valuable 
to overcome the difficulties of integrating a second-order nonlinear differential equation.

The Hamiltonian associated to the Lagrangian of Eq. (\ref{Lag3D}) is given by
\begin{align*}
 & H_{\ell_{\textrm{max}}} \eq  \dot{R}\,\frac{\partial \lambda_{ \ell_{\textrm{max}}} }{\partial \dot{R}}  \me \lambda_{ \ell_{\textrm{max}}} \\
 & \eq   \frac{ [3\,R^2\,\dot{R} \me (14 / {\rm Ca})] \,\sqrt{3\,R^2\,\dot{R} \ma (7 / {\rm Ca})} \me (20 / {\rm Ca}^{3/2})}{(27 / \sqrt{{\rm Ca}}) R^3} \,.
\end{align*}
This Hamiltonian might be a constant of motion which, for convenience, we denote by $(d_{1} \sqrt{{\rm Ca}})/9$, leading us to the following relation
\begin{eqnarray}\label{EqMotion3D}
&[3\,R^2\,\dot{R} \me (14 / {\rm Ca})] \,\sqrt{3\,R^2\,\dot{R} \ma (7 / {\rm Ca})} \me (20 / {\rm Ca}^{3/2})= \nonumber \\
&3\,R^3\,d_1 \,.
\end{eqnarray}
The latter identity should be seen as a first order differential equation for $R(t)$. Such nonlinear differential equation can be made simpler by means 
of defining the function $g(t) \eq [R(t)]^3$, in terms of which Eq. (\ref{EqMotion3D}) is written as
\begin{eqnarray}\label{EqMotion3D-2}
  [ \dot{g} \me (14 /{\rm Ca}) ]\,\sqrt{\dot{g} \ma ( 7 / {\rm Ca}) } \me (20 / {\rm Ca}^{3/2}) \eq 3\, d_1 \, g \,. \nonumber \\
\end{eqnarray}
Now, differentiating the above equation with respect to $t$ and defining $G(t) \equiv \sqrt{\dot{g} + ( 7 / {\rm Ca})}$, we eventually arrive at the following simple relation
\begin{equation}\label{EqMotion3D-3}
  [ G^2 \me ( 7 / {\rm Ca})] (\dot{G} \me d_1) \eq 0 \,,
\end{equation}
whose solutions are $G = \pm \sqrt{( 7 / {\rm Ca})}$ and $\dot{G} = d_1$. The first pair of solutions implies that $g(t)$ is constant and this, in turn, requires $R(t)$ to be constant, which is a non-dynamical solution. Therefore, we conclude that the solution that we are looking for is
$$ G(t) \eq d_1 \,t \ma d_2 \,,$$
where $d_1$ and $d_2$ are arbitrary constants. Then, from the definition of $G(t)$, we have
\begin{align}
 \sqrt{\dot{g} + ( 7 / {\rm Ca})} \eq d_1 \,t \ma d_2 \quad \Rightarrow \quad  \nonumber \\
 g(t) \eq \frac{1}{3}\,d_1^{\,2}\,t^3 \ma d_1\,d_2\,t^2 \ma [d_2^{\,2}\me ( 7 / {\rm Ca})] t \ma d_3 \,, \label{g1}
\end{align}
where $d_3$ is a constant of integration. Nevertheless, the latter expression for $g(t)$ generally is not a solution for Eq. (\ref{EqMotion3D-2}), since the solution for $G(t)$ that gave rise to (\ref{g1}) has been obtained by means of taking the time derivative of the differential equation (\ref{EqMotion3D-2}). In other words, any solution for Eq. (\ref{EqMotion3D-2}) must be of the form (\ref{g1}), but the converse is not true. Indeed, inserting Eq.~(\ref{g1}) into Eq.~(\ref{EqMotion3D-2}), we verify that $d_3$ must be related to $d_1$ and $d_2$ as follows
$$ d_3 \eq \frac{1}{3\,d_1} \left [d_2 \me (5 / \sqrt{{\rm Ca}}) \right ]  \left [d_2 \ma (4 / \sqrt{{\rm Ca}}) \right ] \left [d_2 \ma (1 / \sqrt{{\rm Ca}}) \right ],$$
 in which case the solution $g(t)$ is written as
\begin{widetext}
\begin{equation}
\label{g2}
 g(t) =\frac{1}{3\,d_1} \left [d_1\,t \ma d_2 \me (5 / \sqrt{{\rm Ca}}) \right ] \left [d_1\,t \ma d_2 \ma (4 / \sqrt{{\rm Ca}}) \right] \left [d_1\,t \ma d_2 \ma (1 / \sqrt{{\rm Ca}}) \right] \,.
\end{equation}
\end{widetext}
Finally, since $g = R^3$, we conclude that
\begin{widetext}
\begin{equation}
\label{R3D}
 R(t) = \left \{ \frac{ \left [d_1 t + d_2 - (5 / \sqrt{{\rm Ca}}) \right ] \left [d_1 t + d_2 + (4 / \sqrt{{\rm Ca}}) \right ] \left [d_1 t + d_2 + (1 / \sqrt{{\rm Ca}}) \right ]}{3\,d_1} \right \}^{1/3}  ,
\end{equation}
\end{widetext}
where the constants $d_1$ and $d_2$  can be fixed by using the boundary conditions $R(0)$ and $R(t_f)$. Note that if we take the large capillary number limit $({\rm Ca} \gg 1)$ in the latter solution, we find that $d_{1}=\sqrt{3}\, \left[\,1 - R_{0}\,\right]^{3/2}$, $d_{2}=\sqrt{3}\, R_{0}\, \left[\,1 - R_{0}\,\right]^{1/2}$, and conclude that $R(t)$ becomes a linear function of the time, in accordance with Ref. \cite{solo}. Therefore, the solution obtained in Ref.~\cite{solo} corresponds just the zero order term of the series expansion of the full solution (\ref{R3D}) in powers of the parameter $(1 / {\rm Ca})$.   

\begin{figure}
\includegraphics[width=3.0 in]{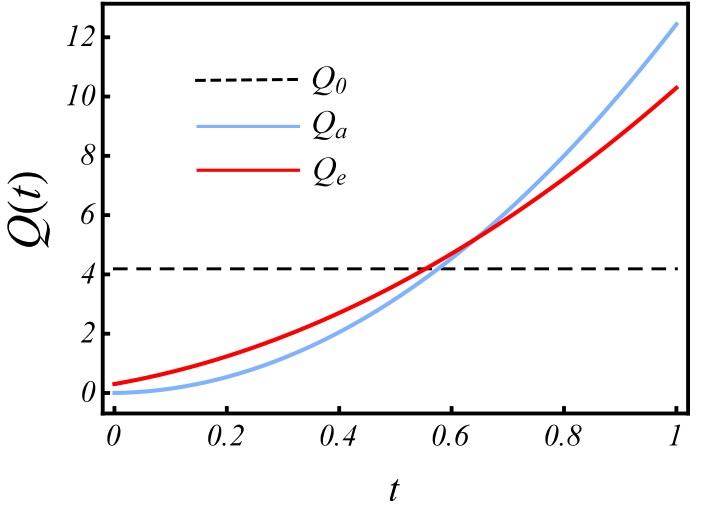}
\caption{(Color online) Injection rate as a function of time for the exact injection rate $Q_{e}$ given by Eq.~(\ref{Q3D}), for the approximate injection rate 
$Q_{a}$ obtained in Ref.~\cite{solo} [see their Eq. (23)], and for the constant injection rate $Q_{0}$ given by Eq.~(\ref{Q03D}).}
\label{injection2}
\end{figure}

Inserting solution (\ref{R3D}) into Eq.~(\ref{newQ}), we obtain that the ideal injection rate for minimizing the fingering is given by the following remarkably simple expression
\begin{equation}\label{Q3D}
  Q(t) \eq \frac{4 \,\pi}{3}\, \left[ \, (d_1\,t \ma d_2 )^2 \me (7 / {\rm Ca})  \,\right] \,.
\end{equation}
Equation~(\ref{Q3D}) is another key result of our current work, offering an exact solution for the injection rate that minimizes the development of fingering instabilities in porous media. This expression reproduces the approximate result obtained in Ref. \cite{solo} [see its Eq. (23)] in the large capillary number limit. In what follows, we will compare the improvement on the minimization of the finger formation attained by the use of the ideal injection rate (\ref{Q3D}) relative to the use of the approximate injection rate obtained in Ref. \cite{solo}.

\begin{figure}
\includegraphics[width=3.2 in]{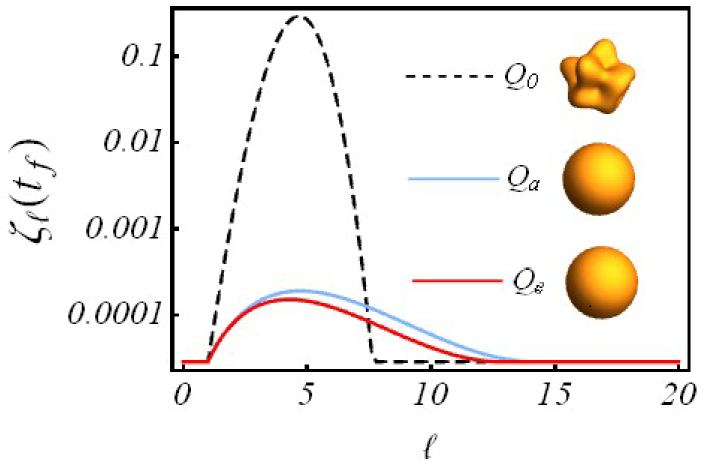}
\caption{(Color online) Log-linear plot of the interfacial perturbation amplitudes $\zeta_{\ell}(t)$ [as given by~Eq.~(\ref{dZeta3D})], as a function
of mode $\ell$ at $t=t_f$, when the three injection rates are used: $Q_0$, $Q_a(t)$ and $Q_e(t)$. The resulting 3D interfaces generated by utilizing 
each of these three injection strategies are also presented.} 
\label{bands2}
\end{figure}

\subsection{Comparing the exact Hamiltonian solution with the approximate Lagrangian solution}
\label{porous3}

Usually, viscous flow in 3D homogeneous porous media is performed by means of a constant injection 
rate $Q_{0}$. By utilizing the dimensionless version of Eq.~(\ref{R2}), so that $R_{f}=1$ and $t_{f}=1$, one easily gets that
\begin{equation}
\label{Q03D}
Q_{0}=\frac{4 \pi}{3}(1 - R_{0}^{3}).
\end{equation}
In this section, we begin our discussion by displaying how the three injection schemes of interest $Q_{e}$, $Q_{a}$, and $Q_{0}$ behave as time progresses. 
This is done in Fig.~\ref{injection2}, that illustrates the time evolution of the exact injection rate $Q_{e}$ given by Eq.~(\ref{Q3D}), as well as the time 
evolution of the approximate injection rate $Q_{a}$, originally obtained in Ref.~\cite{solo} [their Eq. (23)]. Here we set ${\rm Ca}=250$, $R_{0}=0.01$, and 
$\zeta_{n}(0)=R_{0}/350$. The dashed horizontal line represents the constant injection rate $Q_{0}$ obtained from Eq.~(\ref{Q03D}). Despite their common quadratic dependence 
with time, it is clear that the curves representing $Q_{e}$ and $Q_{a}$ behave differently: initially, the $Q_{a}$ curve lies below the $Q_{e}$ curve, and 
subsequently their relative positions are interchanged. These basic differences will result in distinct responses regarding the minimization of the interfacial 
amplitudes, just as in the 2D Hele-Shaw case. It should be recalled that the total volume of injected fluid in the interval $0 \le t \le t_{f}$ 
(given by the area under the curves in Fig.~\ref{injection2}) are actually the same for all pumping rates $Q_{e}$, $Q_{a}$, and $Q_{0}$.

The effect of the different injection rate protocols on the interfacial perturbation amplitudes is depicted in Fig.~\ref{bands2}. It plots the perturbation amplitudes 
given by~Eq.~(\ref{dZeta3D}) at $t_f$, for the approximate injection rate case $\zeta_{\ell}^{a}(t_f)$, the exact ideal pumping rate situation $\zeta_{\ell}^{e}(t_f)$, 
and for the constant injection rate case $\zeta_{\ell}^{0}(t_f)$, as functions of the mode $\ell$. Notice that while using the notation $\zeta_{\ell}^{a}(t_f)$, 
$\zeta_{\ell}^{e}(t_f)$, and $\zeta_{\ell}^{0}(t_f)$, we omitted the subscript $m$ denoting the azimuthal mode number, inasmuch as these perturbation amplitudes 
are spherically symmetric and, therefore, do not depend on $m$. As in Fig.~\ref{injection2}, in Fig.~\ref{bands2} we take ${\rm Ca}=250$, and $R_{0}=0.01$. 
By examining Fig.~\ref{bands2}, one immediately realizes that the amplitudes reached during the constant injection process (dashed curve) are dramatically larger 
than the ones obtained by the exact ($Q_{e}$) and approximate ($Q_{a}$) injection schemes. Moreover, by comparing the behavior of the amplitudes $\zeta_{\ell}^{e}(t_f)$ 
and $\zeta_{\ell}^{a}(t_f)$, one concludes that the exact solution $Q_{e}$ does provide a better minimization of the interfacial amplitudes. 

On the right side of Fig.~\ref{bands2}, we use the amplitudes presented on the left side of it, to plot the shapes of the final 3D fluid-fluid interfaces for 
the injection rates $Q_0$, $Q_a(t)$ and $Q_e(t)$. These linear simulations for the interfaces are obtained by considering the same set of initial conditions, 
and the same random phases for each mode $\ell$. In these particular simulations we have used 18 Fourier modes. It is quite clear that the injection 
schemes related to $Q_a(t)$ and $Q_e(t)$ do restrain the development of interfacial disturbances. However, as pointed out above during the discussion of 
the amplitudes plotted on the left side of Fig.~\ref{bands2}, the exact solution $Q_{e}$ is the one that leads to improved interfacial amplitude minimization.

\begin{figure}
\includegraphics[width=3.2 in]{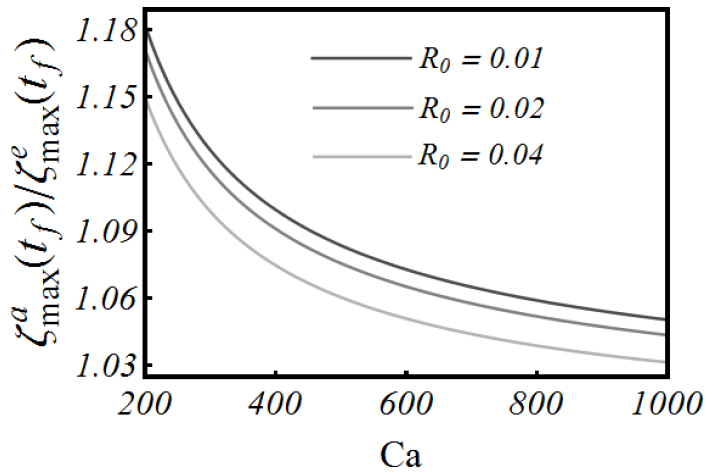}
\caption{Amplitude ratio $\zeta^a_{{\rm max}}(t_f)/\zeta^e_{{\rm max}}(t_f)$ as a function of
${\rm Ca}$, for $R_0=0.01, 0.02$, and 0.04. Here  $\zeta^a_{{\rm max}}(t_f)$ [$\zeta^e_{{\rm max}}(t_f)$]
denotes the maximum amplitude for approximate [exact] injection at $t=t_f$.} \label{sims2}
\end{figure}

We close our current discussion by analyzing Fig.~\ref{sims2}. It plots the maximum amplitude for the approximate
pumping situation divided by the maximum amplitude calculated by using the exact injection rate [$\zeta^a_{{\rm max}}(t_f)/\zeta^e_{{\rm max}}(t_f)$], 
as a function of the capillary number ${\rm Ca}$, at final time $t=t_f$. This is done for three values of initial radius $R_0$: $0.01$, $0.02$, and $0.04$. By 
inspecting Fig.~\ref{sims2}, it is apparent that the ratio $\zeta^a_{{\rm max}}(t_f)/\zeta^e_{{\rm max}}(t_f)$ falls off when ${\rm Ca}$ is increased, and when 
larger values of $R_0$ are used. This is consistent with the approximation considered in Ref.~\cite{solo}, which is valid for large ${\rm Ca}$ and $R_{0}$. It is reassuring 
to observe that the amplitudes of the perturbations for the exact solution $Q=Q_e(t)$ are indeed smaller than the ones obtained by the approximate injection 
solution $Q=Q_a(t)$. By the way, from Fig.~\ref{sims2} one can see that for small values of ${\rm Ca}$ and $R_{0}$, $Q_a(t)$ can lead to a final perturbation 
amplitude $16 \%$ larger than the one obtained through the exact injection process offered by $Q_e(t)$. So, as in the effectively 2D Hele-Shaw case studied 
in Sec.~\ref{HSC3}, our exact injection rate solution~(\ref{Q3D}) is in fact more efficient than the approximate solution $Q_a(t)$ (calculated in Ref.~\cite{solo})
in providing minimization of interfacial disturbances for viscous flow in a 3D homogeneous porous medium.

\section{Concluding remarks}
\label{conclude}

In this paper, we have studied the minimization process of the linear interfacial perturbation amplitudes which arise during the development of viscous fluid fingering phenomena in effectively 2D Hele-Shaw cells, as well as in 3D homogeneous porous media. By employing a Hamiltonian formalism, we have been able to find analytically the exact functional forms for the time-dependent injection rates $Q(t)$ that lead to minimal interfacial deformation. The advantage of using the Hamiltonian approach for the variational problem is that we have to integrate a first-order differential equation instead of a second-order one, something that greatly simplifies the attainment of the analytical solution. In contrast to previous investigations that found approximate solutions for $Q(t)$, our exact results are valid for arbitrary values of the capillary number. Comparison of these exact solutions with their correspondent approximate counterparts reveals that our current results promote a more efficient minimization of the perturbation amplitudes. This improvement is particularly significant for small values of the capillary number.

Our present theoretical work makes specific linear stability predictions that have not yet been subjected to either experimental or nonlinear numerical check of the interface dynamics. Hopefully, a fully nonlinear study could reveal that the exact injection rate provides an even greater improvement on the fingering minimization (in comparison with the approximate solution) than the one we observed here by means of a linear analysis. In this sense, we hope our study will instigate further theoretical and experimental investigations on this rich and challenging research topic.

\begin{acknowledgments}
J.A.M. thanks CNPq for financial support. E. O. D. acknowledges financial support from FACEPE through PPP Project No. APQ-0800-1.05/14.
\end{acknowledgments}

\end{document}